\documentclass[aps,epsfig]{revtex4}
\usepackage{graphicx}
\usepackage{epstopdf}
\usepackage{latexsym}
\usepackage{amssymb}
\usepackage{textcomp}
\usepackage{amsmath}
\usepackage[center]{subfigure}
\usepackage{feynmp}
\usepackage{slashed}
\usepackage{pdfpages}
\begin{document}
\renewcommand\thesection{\arabic{section}}
\renewcommand\thesubsection{\thesection.\arabic{subsection}}
\renewcommand\thesubsubsection{\thesubsection.\arabic{subsubsection}}
 \newcommand{\bq}{\begin{equation}}
 \newcommand{\eq}{\end{equation}}
 \newcommand{\bqn}{\begin{eqnarray}}
 \newcommand{\eqn}{\end{eqnarray}}
 \newcommand{\nb}{\nonumber}
 \newcommand{\lb}{\label}
\title{A Macroscopically Effective Lorentz Gauge Theory of Gravity}
\author{Ahmad Borzou}
\email{ahmad_borzou@baylor.edu}
\unitlength 0.5mm
\affiliation{Physics Department, Baylor
University, Waco, TX 76798-7316, USA}

\date{\today}

\begin{abstract}

Following the ideas of effective field theories, we derive classically effective field equations of recently developed Lorentz gauge theory of gravity. It is shown that Newton's gravitational constant emerges as an effective coupling parameter if an extremely small length is integrated out of the underlying theory. The linear version of the effective theory is shown to be fully consistent with the Newtonian gravity. We also derive a numerical solution for the interior of a star and show that in the non-linear regions, the behavior of the effective theory deviates from the predictions of general relativity. 
 
\end{abstract}

\maketitle

\section{Introduction}
The standard model, which is until now the best known description of the fundamental fields, is a gauge theory of the group 
$SU(3)\times SU(2) \times U(1)$, a geometry of internal spaces over the space-time, and except gravity explains all the forces of the nature. In the standard model the connections, known as the gauge bosons, are the dynamical variables and as a result their Lagrangians are polynomials of order four and consequently their theories are successfully renormalizable. General relativity on the other hand is a Poincare gauge theory in which the metric, and not the connections, is the dynamical variable.
If one expands the metric around a classical background, a polynomial of infinite orders will be reached and as one goes to higher orders in the perturbation, the theory becomes more and more divergent. A review of these ultraviolet divergences can be found in \cite{Weinberg:1980gg}. It is therefore legitimate to search for a gauge formulation of gravity based on geometry of internal spaces over space-time. This avenue of research was opened by the works of Utiyama, Sciama, and Kibble \cite{PhysRev.101.1597,RevModPhys.36.463, Kibble1961} and will be discussed more in the next section. There has been a glaring progress in this area of research but gravity has not been incorporated into the standard model until now.
In \cite{Borzou2016} we have introduced a framework in which the metric remains non-dynamical. In this work, it is shown that the degree of divergence does not increase if one tries to calculate higher and higher orders of perturbation. Therefore, the theory remains predictable at high energies. Although the same formulation should in principle work in the macroscopic scales as well, it is extremely more convenient if a macroscopically effective version of the theory can be found. 

Fortunately the astonishing feature of the nature is that at any scale, one can find an ``effective'' physics by isolating the interesting phenomena of that scale without worrying about the fundamental rules behind it. The description of the important physics of a given scale is referred to as the effective theory of that scale and can be found if there exist some parameters in the underlying theory that are very large or very small in comparison with the quantities of interest in that scale. It is most of the time possible to absorb the very large or very small parameters into some other constants of the theory and then set them to infinity or zero afterward. A detailed description of the ideas in effective theories can be found in \cite{Burgess:2007,Bain:2013imf,GEORGI1993}. Fermi theory of beta decay, as an example, is an effective theory of the electroweak interactions and is built by absorbing the mass of the W boson, which is very large in the MeV scale, into the coupling constant of the electroweak theory and defining the effective Fermi constant
\bqn
\lb{FermiConstant}
G_{\text{F}}=\frac{\sqrt{2}}{8}\frac{g^2}{m^2_{\text{W}}},\nb
\eqn
where $g$ is the coupling constant of the electroweak theory and $m_{\text{W}}=80$ GeV is the mass of the W boson and is set to infinity in the effective theory. 
In this paper we would like to follow the same ideas and find an effective Lorentz gauge theory in a general non-linear space-time. We first note that if the linear version of a given theory is known and the equivalence principle is valid,
the non-linear format of the theory is uniquely given by the principle of minimal coupling \cite{Will2006}. Therefore, all we need is to find the linear version of the effective theory and then simply use the minimal coupling principle and convert it to the full non-linear effective theory. 

The present paper is organized as follows. A brief review of Lorentz gauge theory of gravity is presented in section 2. In section 3 the linear version of effective Lorentz gauge theory is derived. The general non-linear effective theory is introduced in section 4 where the interior of an incompressible star is studied. A conclusion is drawn at the end in section 5.

\section{Lorentz Gauge Theory of Gravity}
\lb{reviewOfLGT}
The global Poincare symmetry is thus far preserved in all the interactions that are observed by the experiments in the particle physics and therefore, inspired by Yang-Mills theories, one may wish to locally preserve the Poincare invariance. This requires adding a class of gravitational compensating fields, connections, to preserve the local translations. Here in contrast to general relativity where only translations are considered, one can in addition define a class of inertial frames at each point of space-time that are related to each other by the Lorentz transformations. In order to locally preserve the latter, one also needs to insert a second type of compensating fields called the spin connections. 
There is a wide agreement on the kinematics of the gauge theories of gravity that was mentioned above. However, the dynamics, i.e. how matter influences the geometry, is still an active area of research. There are a large number of invariant scalars constructed out of the curvature tensor and one needs to seek extra conditions in order to reduce the possibilities.
Therefore, in contrast to Yang-Mills theories, the Lagrangian is not uniquely determined by setting the symmetry of the theory. 
We do not intend to review the gauge theories of gravity in the present article but only refer to some excellent papers devoted to the subject \cite{Hehl19951, Blagojevic:2013xpa,IVANENKO19831,RevModPhys.48.393,Obukhov:2006gea,Shapiro2002113}. The list is far from being complete and more details can be found in literature.
There is also another important difference between the Yang-Mills theories and the Poincare gauge theories of gravity. In the former the internal symmetries are defined on the space-time while the latter describe the geometry of space-time itself with the metric being a dynamical variable. Consequently, energy-momentum tensor is the source of gravity in the Poincare gauge theories which itself leads to an infinite number of fundamentally independent interactions \cite{Hamber2009} comparing to a few interactions in the Yang-Mills theories. It is hard to imagine the renormalizability of a theory with infinite number of fundamental interactions. This however is not the only hurdle in quantizing a theory whose dynamical variable is the metric. Quantizing the space-time itself is in a sharp contrast with the current picture of quantum mechanics where fields are quantized on the space-time and also where time has a very specific external role. This role does not conceptually fit in a theory of gravity where time itself is going to be quantized, i.e. is dynamical \cite{Isham1991,kiefer2012}. Alternatively, one may wish to find a formalism of gravity in which the metric is not dynamical. Lorentz gauge theory is an attempt in this direction and is briefly reviewed in this section. For more details one can consult \cite{Borzou2016}. 

The equivalence principle implies that there is a free falling frame at any point in
space-time which is both a Lorentz and a 
coordinate frame. Therefore, tetrad field can be divided into a part which contains the angle between the free falling frame and that associated with the arbitrarily chosen coordinates and the part that contains the
angle between the free falling frame and the arbitrarily chosen Lorentz frame. Mathematically this means
\bq
e_{i \mu}=\eta^{\bar{k} \bar{l}} e_{i \bar{k}}e_{\bar{l}\mu},
\eq
 where the free falling frame is shown with a bar while the Latin indices indicate the Lorentz frame and the Greek ones refer to
 the coordinate system. Either of the two tetrad constituents can be dynamical. Namely,
\bqn
\lb{TwoVariCases}
\delta e_{i \mu}=
\begin{cases}
\eta^{\bar{k} \bar{l}}e_{i \bar{k}}\delta e_{\bar{l} \mu} &  \text{Case I},\\
\eta^{\bar{k} \bar{l}}\delta e_{i \bar{k}}e_{\bar{l} \mu} &  \text{Case II}.
\end{cases}
\eqn
General relativity and the Poincare gauge theories are based on the first case. Lorentz gauge theory is built upon the second one and develops
no dynamics in the metric because 
\bq
g_{\mu \nu}=\eta^{ij}e_{i \mu}e_{j \nu}
=\eta^{\bar{i}\bar{j}}e_{\bar{i} \mu}e_{\bar{j} \nu}.
\eq
Therefore, 
\bqn
\lb{deltag}
\delta g_{\mu \nu}=0,
\eqn
in the latter case and as a result variation of whatever made of the metric
including the Christoffel symbols and the determinant of tetrad is zero
\bqn
&&\delta \Gamma^{\alpha}_{\mu \nu}=0\nb\\
&&\delta e=\delta \sqrt{-g}=0.\nb
\eqn
Moreover, variation of the spin connections for the second case is
\bqn
\lb{VariationOfA}
\delta A_{ij\mu}=D_{\mu}\big(e_j^{~\nu}\delta e_{i \nu}\big),
\eqn
which is different from that in the first case by not having one additional term containing $\delta \Gamma^{\alpha}_{\mu\nu}$ \cite{Borzou2016}.
From here it is easy to see that even for an identical Lagrangian, the field equations will be quite different for the two cases above. 
The Lorentz gauge theory is defined by the following action
\bqn
\lb{action}
S&=&\int e d^4x\Big[{\cal{L}}_{M}+{\cal{L}}_{A}\Big],
\eqn
where $e$ is the determinant of the tetrad field, ${\cal{L}}_{M}$ is the Lagrangian of matter and ${\cal{L}}_{A}$ is the gauge field's Lagrangian to be determined later. We first would like to mention that in principle ${\cal{L}}_{M}$ can be the Lagrangian of any field including scalars, vector bosons, and fermions. However, since the Lorentz gauge theory is constructed upon the second case in equation (\ref{TwoVariCases}) where the variation of the metric as well as its determinant are zero, the bosonic fields do not enter the field equations.
This can be understood by investigating the Lagrangian of electrodynamics ${\cal{L}}_{\text{Elec}}=-\frac{1}{4}F_{\mu \nu}F^{\mu \nu}$ as an  example.
To find its contribution to the field equations, we need to evaluate the following
\bqn
\frac{{\delta\big(\sqrt{-g}\cal{L}}_{\text{Elec}} \big)}{\delta g_{\mu \nu}}\delta g_{\mu \nu}+
\frac{{\delta\big(\sqrt{-g}\cal{L}}_{\text{Elec}} \big)}{\delta e_{i \mu}}\delta e_{i\mu},
\eqn
where $\delta g_{\mu \nu}$ and $\delta e_{i\mu}$ are themselves functions of $\delta e_{i\bar{k}}$ that is introduced in the second case in equation \eqref{TwoVariCases}. This function for $\delta g_{\mu \nu}$ is zero according to equation \eqref{deltag} and therefore the first term in the latter equation is zero. On the other hand the second term is also zero because the tetrad field has not directly contributed here, but only through the metric. It is worth mentioning that the second term is not zero for fermionic Lagrangians where the tetrad is coupled to the Dirac matrices. 
Although this statement is in sharp contrast with general relativity, there is not only no experimental evidence that bosonic fields gravitate but also some are suggesting otherwise. Cosmological observations are pointing that the vacuum energy, that is confirmed to be genuine \cite{Casimir1948,Sparnaay1957}, is not generating the field that is predicted by general relativity and is hundreds of orders of magnitude larger than what is being observed \cite{RevModPhys.61.1}. This can be easily explained by our formalism because the vacuum energy is being described by ${\cal{L}}_{_{\text{Vac}}}=\rho_{_{\text{Vac}}}=\text{const.}$ whose variation is zero for the same reasons described above and therefore does not contribute to the field equations. On the other hand, in \cite{Borzou2016} we have shown that when the matter content in a homogeneous and isotropic universe is negligible, the de Sitter space-time is an exact solution of the theory and therefore the positive rate expansion of the universe at the present time can be explained with no need for a dark energy. 

The matter field is therefore limited to fermions including charged and neutral leptons as well as quarks. The same argument indicates that within the framework of Lorentz gauge theory of gravity, dark matter is a fermion. Although the nature of the dark particles is not yet known, many of the candidates are of fermionic nature \cite{Baltz:2004tj}.

In order to derive the field equations, one needs to write $\delta A_{ij\mu}$ in terms of $\delta e_{i \mu}$ or vice versa since they are dependent according to equation (\ref{VariationOfA}). We first try to eliminate $\delta A_{ij\mu}$ in terms of $\delta e_{i \mu}$ and write the field equation as 
\bq
\lb{tetradfieldeq}
\frac{\delta(e {\cal{L}}_A)}{\delta e_{i\mu}}=
- \frac{\delta(e {\cal{L}}_M)}{\delta e_{i\mu}}.
\eq
However as is shown in \cite{Borzou2016} the right hand side, the source term, will be zero and hence the tetrad is a non-propagating field. We are then forced to eliminate $\delta e_{i \mu}$ in terms of $\delta A_{ij\mu}$. This can be achieved by means of the Lagrange multiplier method. A constraint Lagrangian will be added to the action and then the two fields $A_{ij\mu}$ and $e_{i \mu}$ will be treated independently which is the usual process in the method. The action therefore is 
\bqn
\lb{action}
S&=&\int e d^4x\Big[{\cal{L}}_{M}+{\cal{L}}_{C}+{\cal{L}}_{A}\Big],
\eqn
where ${\cal{L}}_{C}$ is just the tetrad postulate, $D_{\mu}e_{i \nu}=0$, inserted as a constraint 
\bq
\lb{LC}
{\cal{L}}_{C}=S^{\mu \nu i}D_{\mu}e_{i \nu},
\eq
with $S^{\mu \nu i}$ being a Lagrange multiplier. 

Finally, ${\cal{L}}_{A}$, is the gauge field's Lagrangian and its most general form is \cite{Hayashi01091980, PhysRevD.80.104031}
\bqn
\lb{LA}
{\cal{L}}_{A}&=&-\frac{1}{4}\Big(c_1 F_{\mu\nu ij}e^{i \mu}e^{j \nu} + c_2 F_{\mu\nu ij}F^{\mu \sigma ik}e^{j \nu}e_{k \sigma}
+c_3 F_{\sigma \nu mj}F_{\mu \alpha in}e^{j \nu}e^{i \mu}e^{m \sigma}e^{n \alpha}\nb \\
&&~~~+c_4F_{\mu \nu ij}F^{\alpha \beta mn}e^{i \mu}e^j_{~\beta}e_{m \alpha}e_n^{~\nu}+c_5F_{\mu \nu ij}F^{\mu\nu ij}\Big),
\eqn
where
\bq
\lb{strength}
F_{\mu\nu ij}=\partial_{\nu}A_{ij\mu}-\partial_{\mu}A_{ij\nu}+A_{i~~\mu}^{~m}A_{mj\nu}-A_{i~~\nu}^{~m}A_{mj\mu}.
\eq
Here $A_{ij\mu}$ are the gauge fields, the spin connections, and are related to the tetrad field through the tetrad postulate
\bq
\lb{SpinCon}
A_{ij \mu}=e_j^{~\nu}\partial_{\mu}e_{i \nu}-\Gamma^{\alpha}_{\mu \nu}e_{i \alpha}e_j^{~\nu},
\eq
where $\Gamma^{\alpha}_{\mu \nu}$ are the metric compatible Christoffel symbols
\bq
\lb{Chris}
\Gamma^{\alpha}_{\mu \nu}=\frac{1}{2}g^{\alpha \beta}(\partial_{\nu}g_{\mu \beta}+\partial_{\mu}g_{\nu \beta}
-\partial_{\beta}g_{\mu\nu}).
\eq
To get the closest equations to those of the standard model, one needs to abandon the first four terms in the Lagrangian (\ref{LA}). Except the first term that leads to a non-propagating field equation, the terms are dropped because they insert the strength tensor, $F_{\mu \nu ij}$, into the source field \cite{Borzou2016} which is something we haven't seen in the standard model. The Lagrangian can be set to
\bqn
\lb{NLA}
{\cal{L}}_{A}&=&-\frac{1}{4}c_5F_{\mu \nu ij}F^{\mu\nu ij}.
\eqn

The filed equations corresponding with this Lagrangian and the second case in \eqref{TwoVariCases} read
\bqn
\lb{fieldeq}
&&\frac{\delta{\cal{L}}_M}{\delta e_{i \nu}}-D_{\mu}S^{\mu \nu i}=0,\nb\\
&&c_5D_{\nu}F^{\mu \nu ij}=\frac{\delta{\cal{L}}_{M}}{\delta A_{ji \mu}}+S^{\mu \nu [i}e^{j]}_{~\nu}.
\eqn
To solve the first equation, we suppose 
\bqn
\lb{LagrangeMultiplier}
S^{\mu \nu i}=T^{\mu \nu}\xi^i,
\eqn 
where the energy-momentum tensor is defined as $T^{\mu \nu} \equiv e_i^{~\mu}\frac{\delta{\cal{L}}_M}{\delta e_{i \nu}}$.
Substituting this into the equation and using the conservation law of energy-momentum, the equation reads
\bqn
T^{\mu\nu}\left(e^i_{~\mu}-D_{\mu}\xi^i\right)=0.
\eqn 
Hence, the proposed form for $S^{\mu \nu i}$ is a solution if  
\bqn
\lb{DXiEquation}
D_{\mu}\xi^i=e^i_{~\mu} ~~\text{where}~~ T^{\mu\nu} \neq 0.
\eqn 
It should be noted that the field equations do not specify the value of $\xi^i$ where matter is absent, i.e. where $T^{\mu \nu}=0$. However, the vector is always multiplied by the energy-momentum tensor which is zero exactly where the vector is unknown. To find out $\xi^i$ inside matter, it should be noted that any matter distribution is a collection of units of matter. So, all we need is to find the solution to $\xi^i$ for a single unit of matter. We assume a unit of matter is defined within a sphere of radius $\delta$ that will be set to zero later. The solution therefore is 
\bqn
\lb{multiplier}
\xi^i(x)=
\lb{xiDef}
\begin{cases}
e^i_{~\alpha}(X)(x^{\alpha}-X^{\alpha}) & x < \delta, \\
\xi^i_{_{+}} & x \geq \delta, 
\end{cases}
\eqn 
where $X$ refers to a local point, and $\xi^i_{_{+}}$ will be multiplied by $T^{\mu\nu}=0$ and will drop out of the equations. The solution holds only in the free falling frame co-moving with the unit of matter. The solution in an arbitrary frame can be found by a simple frame transformation.  
An interpretation of equations above is that a matter distribution can be divided to extremely small cells, spheres of radius $\delta$. Inside each cell we can define a local position vector $\vec{\xi}$ which is defined with respect to the center of that cell. A pictorial illustration is given in figure (\ref{CellsDef}).

Equation (\ref{fieldeq}) together with (\ref{LagrangeMultiplier}) and (\ref{multiplier}) lead to the final form of field equations
\bqn
\lb{SimpleField}
c_5D^{\nu}F_{\mu\nu ij}&=&J_{\mu ij}\nb\\
&=&\frac{\delta{\cal{L}}_{M}}{\delta A^{ji \mu}}+\frac{1}{2}T_{\mu j}\xi_i-\frac{1}{2}T_{\mu i}\xi_j.
\eqn

In the end we would like to mention that 
Lagrangian \eqref{NLA} is not new at all. In literature this is known as the curvature squared Lagrangian and within the context of gauge theories of gravity, it was first proposed by von der Heyde \cite{VonDerHeyde1975} and further studied by
Hehl, Nietsch, Blagojevic, Nikolic, Dimakis, Mielke, Minkevich, and others, for example see \cite{ASNA:ASNA2103020310,Obukhov1989,BAEKLER1987800,Vereshchagin:1999jv,Kuhfuss1986}. In the early versions of the Poincare gauge theories, the Lagrangian was linear in the curvature, defining the so called Einstein-Cartan theory. However, people realized that such a Lagrangian will lead to non-propagating contact type interactions. These type of interactions can only be interpreted as the emergences of more fundamental ones. Therefore, it was essential to introduce the curvature squared Lagrangians in order to reap a propagating field. Although
Lagrangian \eqref{NLA} is extensively used
within the Poincare gauge theories, their field equations are to some extent different than those in equation \eqref{SimpleField}. This again roots back to the fact that the dynamical variable in the Poincare gauge theories corresponds with the first case in equation \eqref{TwoVariCases} while in Lorentz gauge theory corresponds with the second one.

\section{Effective Lorentz Gauge Theory: Linear Static Case}
\lb{linearClassicalMatter}

\begin{figure}[tbp]
\centering
\includegraphics[width=6cm]{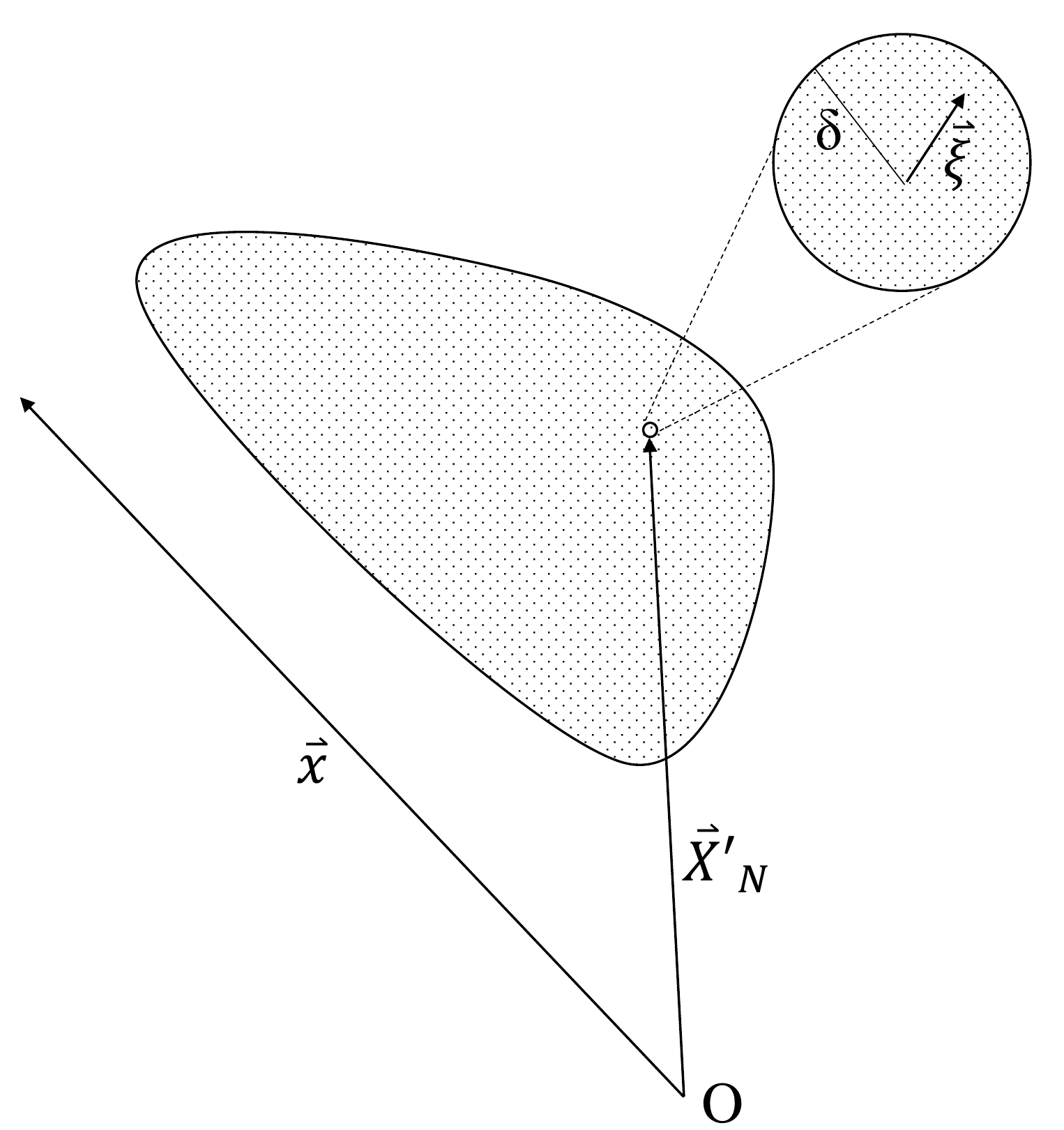}
\caption{A matter distribution represented by an energy-momentum tensor $T_{\mu \nu}$. The distribution is made of N number of tiny cells, spheres of radius $\delta$. The $\vec{\xi}$ vector is defined with respect to the center of each cell while $\vec{X}'_N$ represents the location of the center of the Nth cell in an arbitrary coordinate system. Also, $\vec{x}$ refers to an arbitrary location.}
\lb{CellsDef}
\vspace{0.75cm}
\end{figure} 

This section is devoted to studying linear and static version of Lorentz gauge theory and deriving its corresponding effective theory that holds in macroscopic scales. The starting point is to decompose the metric into the Minkowski metric plus a perturbation 
\bqn
g^{\mu \nu} \equiv \eta^{\mu \nu} + h^{\mu \nu}, ~~\text{where}~~|h^{\mu\nu}| \ll 1.
\eqn
After inserting this in equation (\ref{SimpleField}) and neglecting terms of order two and higher in $h_{\mu\nu}$, the linear field equation reads 
\bqn
&&\partial^2A_{ij \mu}=c_5^{-1}J_{\mu ij},
\eqn
where $\eta^{\alpha \nu}\partial_{\alpha}\partial_{\nu}$ is replaced by $\partial^2$ and the Lorentz gauge is fixed by 
$\eta^{\alpha \nu}\partial_{\alpha}A_{ij \nu}=0$. In a static problem the d'Alembertian is replaced by a Laplacian 
\bqn
\lb{linearOriginal}
\nabla^2A_{ij \mu}=c_5^{-1}J_{\mu ij},
\eqn
and the Green's function is the inverse of the Laplace operator
\bqn
\lb{GreenFunction}
G(\vec{x}-\vec{x}'')=\frac{1}{4\pi |\vec{x}-\vec{x}''|}.
\eqn

Equation \eqref{linearOriginal} and its Green's
function look very much like a simple electrostatic problem. However, there is a disappointing difference. Even for the simplest and highly symmetric matter distributions, the source field $J^{\mu ij}$ fluctuates extremely rapidly due to the nature of $\xi^i$. The directions of this vector can be very different at two adjacent points. The non-uniformity of this vector can be seen if one moves from a unit of matter to the adjacent one and will ruin the symmetries of the problem. Consequently, one needs to take into account the fluctuations of the vector in scales as small as $\delta$ even though he is solely interested in large scale description of the configuration. Although it is still in principle possible to solve the problem, practically it is frustratingly hard. 

In the following we will find an effective theory of the underlying field equations to overcome the undesired fluctuations in the source field $J_{\mu ij}$. The small parameter $\delta$ in equation \eqref{xiDef} is analogues to the inverse of the W boson's mass that was absorbed into the coupling constant of the electroweak theory, something that led to the Fermi theory of beta decay. Therefor, if $\delta$ could be absorbed into the coupling constant of Lorentz gauge theory $c_5^{-1}$, an effective theory would be reached. To investigate the possibility of integrating out $\delta$, we start with the full theory and calculate
the gravitational field of the Nth cell in figure \ref{CellsDef}  
\bqn
\lb{FieldOfACell}
A_{ij\mu_{_N}}=-c_5^{-1}\int\frac{d^3x'' J_{\mu ij}(x'')}{4\pi |\vec{x}-\vec{x}''|},
\eqn
where 
\bqn
\vec{x}~''\equiv\vec{X}_N'+\vec{\xi},
\eqn
as is shown in the figure, and the integral is over the interior of the cell. Therefore, (\ref{FieldOfACell}) can be rewritten as 
\bqn
A_{ij\mu_{_N}}&=&-c_5^{-1}\int\frac{d^3\xi J_{\mu ij}(x'')}{4\pi |\vec{x}-\vec{X}'_N-\vec{\xi}|}\nb\\
&=&-\frac{c_5^{-1}}{8\pi} \int\frac{d^3\xi [T_{\mu j}(x'')\xi_i-T_{\mu i}(x'')\xi_j]}{|\vec{x}-\vec{X}'_N-\vec{\xi}|},
\eqn
where the spin angular momentum, $\frac{\delta{\cal{L}}_{M}}{\delta A^{ji \mu}}$, is neglected as it is not relevant classically and 
\bqn
\xi_{i}= \xi
\begin{pmatrix}
0\\
\sin(\theta_{\xi})\cos(\phi_{\xi})\\
\sin(\theta_{\xi})\sin(\phi_{\xi})\\
\cos(\theta_{\xi})
\end{pmatrix}.
\eqn
The temporal component is zero because in a classical problem the gravitational field inside the cell can be neglected which implies that all the observers co-moving with the cell are simultaneous. 
As far as a classical matter is concerned, the energy-momentum tensor can be assumed constant within the cell and therefore 
\bqn
\lb{linearField1}
A_{ij\mu_{_N}}&=&-\frac{c_5^{-1}}{8\pi} \Big[T_{\mu j}(X'_N)\int \frac{d^3\xi~\xi_i}{|\vec{x}-\vec{X}'_N-\vec{\xi}|}-T_{\mu i}(X'_N)\int\frac{d^3\xi~\xi_j}{|\vec{x}-\vec{X}'_N-\vec{\xi}|}\Big]\nb\\
&=&-\frac{c_5^{-1} \delta ^2}{40\pi}\delta V \Big[
T_{\mu j}(X'_N)\frac{(\vec{x}-\vec{X}'_N)_i}{|\vec{x}-\vec{X}'_N|^3} y_i 
-T_{\mu i}(X'_N)\frac{(\vec{x}-\vec{X}'_N)_j}{|\vec{x}-\vec{X}'_N|^3} y_j
\Big],
\eqn
where $ |\vec{x}-\vec{X}'_N| > \xi $ is assumed and $\delta V = \frac{4\pi}{3}\delta^3$ is the volume of the cell, $y_i$ is equal to one for spatial components and to zero for temporal one, and
\bqn
T_{\mu j}\delta V=\frac{p_{\mu}p_{j}}{E},
\eqn
when $E$ is the energy of the cell and $p_{\mu}$ is its momentum vector. Therefore, the final form of the gravitational field of the Nth cell is 
\bqn
\lb{GravityOfCell}
A_{ij\mu_{_N}}&=&-\frac{c_5^{-1} \delta ^2}{40\pi} \Big[
\frac{p_{\mu}p_{j}}{E}\cdot\frac{(\vec{x}-\vec{X}'_N)_i}{|\vec{x}-\vec{X}'_N|^3} y_i 
-\frac{p_{\mu}p_{i}}{E}\cdot\frac{(\vec{x}-\vec{X}'_N)_j}{|\vec{x}-\vec{X}'_N|^3} y_j
\Big].
\eqn
We would like to find an effective theory that leads to the same gravitational field for regions outside of the cell while fails to explain the gravity inside the cell. To do so the equation can be rewritten as
\bqn
A_{ij\mu_{_N}}&=&-\frac{c_5^{-1} \delta ^2}{40\pi} \int d^3z \delta^3(\vec{z}-\vec{X}'_N) \Big[
\frac{p_{\mu}p_{j}}{E}\cdot\frac{(\vec{x}-\vec{X}'_N)_i}{|\vec{x}-\vec{X}'_N|^3} y_i 
-\frac{p_{\mu}p_{i}}{E}\cdot\frac{(\vec{x}-\vec{X}'_N)_j}{|\vec{x}-\vec{X}'_N|^3} y_j
\Big]\nb\\
&=&-\frac{c_5^{-1} \delta ^2}{40\pi} \int d^3z \delta^3(\vec{z}-\vec{X}'_N) \Big[
\frac{p_{\mu}p_{j}}{E}\cdot\frac{(\vec{x}-\vec{z})_i}{|\vec{x}-\vec{z}|^3} y_i 
-\frac{p_{\mu}p_{i}}{E}\cdot\frac{(\vec{x}-\vec{z})_j}{|\vec{x}-\vec{z}|^3} y_j
\Big]\nb\\
&=&-\frac{c_5^{-1} \delta ^2}{40\pi} \int d^3z \delta^3(\vec{z}-\vec{X}'_N) \Big[
\frac{p_{\mu}p_{j}}{E}\cdot\partial_{z_i}\frac{1}{|\vec{x}-\vec{z}|} y_i 
-\frac{p_{\mu}p_{i}}{E}\cdot\partial_{z_j}\frac{1}{|\vec{x}-\vec{z}|} y_j
\Big]\nb\\
&=&\frac{c_5^{-1} \delta ^2}{40\pi} \int d^3z \Big[
\frac{\partial_{z_i}\left(\frac{p_{\mu}p_{j}}{E}\delta^3(\vec{z}-\vec{X}'_N)y_i\right)}{|\vec{x}-\vec{z}|}  
-\frac{\partial_{z_j}\left(\frac{p_{\mu}p_{i}}{E}\delta^3(\vec{z}-\vec{X}'_N)y_j\right)}{|\vec{x}-\vec{z}|} 
\Big].
\eqn
We should note that $\frac{p_{\mu}p_{i}}{E}\delta^3(\vec{z}-\vec{X}'_N)$ is simply the energy-momentum tensor of a point-like particle \cite{Weinberg1972} and can be replaced by $T_{\mu i}(z)$, i.e. the cell will be treated like a point-like quantity from now on. Taking a Laplacian of both sides results in the effective field equations
\bqn
\lb{linearField5}
\nabla ^2 A_{ij \mu}= 4\pi G \tilde{J}_{\mu i j},
\eqn
where 
\bqn
\lb{linearstaticsource}
\tilde{J}_{\mu i j}=y_j\partial _j T_{\mu i}-y_i\partial _i T_{\mu j},
\eqn
is the macroscopic version of the gravitational source for a static matter distribution and
\bqn
G=\frac{c_5^{-1} \delta ^2}{40 \pi},
\eqn
is Newton's gravitational constant, the effective coupling constant in our case. At this point there are two notes in order. First, the $\delta^2$ in this equation could have been absorbed into the coupling constant of Lagrangian \eqref{NLA} right from the beginning by a simple scaling, $c_5 \rightarrow \delta^2 c_5$. Second, a comparison of the effective coupling constant $G$ and the Fermi constant $G_{\text{F}}$ can be inspiring. The Fermi constant is very small in magnitude in comparison with the coupling constants of the electromagnetic or strong interactions because of the appearance of the square of the W boson's mass in its denominator. Inspired by these historical notes, we can hope that the coupling constant of Lorentz gauge theory $c_5^{-1}$ has a magnitude not very different than those of the other three forces and the Newton's gravitational constant is very small because $c_5^{-1}$ is divided by the square of a huge mass $\frac{1}{\delta}$. 

Before proceeding further, it should be emphasized that a function whose Laplacian is zero can be added to Green's function \eqref{GreenFunction} and can be determined by boundary values. Putting it equal to zero, like what we did above, is equivalent to assuming that the gauge field rolls toward zero at infinity. The extra function is not zero in most of the curvature squared Poincare theories because they have a potential that grows linearly with the distance from a given source \cite{SIJACKI1982435,ASNA:ASNA2103020310}. We would like to examine whether the present theory also contains a potential of this kind and if it does, our choice of Green's function is not correct. To find the most general form of the gravitational field of a single particle in the non-relativistic limit, the appropriate source field $\tilde{J}_{\mu ij}=4\pi G y_i \partial_i \rho$ corresponding with the energy-momentum tensor $T_{\mu i}=-\rho \delta_{\mu 0}\delta_{i 0}$ should be substituted into equation \eqref{linearField5}
\bqn
\nabla ^2 A_{i00}= 4\pi G y_i \partial_i \rho. \nb
\eqn
In a spherically symmetric case, every vector should be radial and therefore $A_{i 00}=\frac{1}{2}a'\hat{r}_i$ and $\partial_i \rho=\rho'\hat{r}_i $ where $a$ is the potential, prime indicates a derivative with respect to the distance from the center and $\hat{r}$ is the radial unit vector in the spherical coordinate system and $i$ refers to the Cartesian indices. Moreover, except the unit vector, every other quantity is only a function of the distance from the center. By a straightforward calculation, this equation can be written in the following form 
\bqn
a'''+\frac{2}{r}a''-\frac{2}{r^2}a'=8\pi G \rho',
\eqn
which is a second order differential equation in terms of $a'$. The most general solution of which is 
\bqn
a'=a'_1 + a'_2+\left(a'_2 \int^r \frac{a'_1 F dr}{W}-a'_1 \int^r \frac{a'_2 F dr}{W}\right),
\eqn
where $a'_1=\frac{b_1}{r^2}$ and $a'_2=b_2 r$ are the solutions to the homogeneous part of the equation, $W=\frac{3}{r^2}$ is the Wronskian of the two, $b_i$ are arbitrary constants, and F is the source $8\pi G \rho'$ while $\rho$ for a single particle is $m \frac{\delta(r)}{4\pi r^2}$ with $m$ being the mass of the particle. 
Substituting these into the latter equation and taking the integral $\int dr a'$ gives the Newtonian potential
\bqn
a=-\frac{b_1}{r}+ \frac{b_2 r^2}{2}-\frac{2Gm}{r}+b_3,
\eqn
where there is no track of the so called confinement potential that linearly increases with the distance from the source and that can't be eliminated by manipulating the integration constants. The fact that, despite starting from the same Lagrangian, the non-relativistic field in the Poincare gauge theories is different than that in Lorentz gauge theory, stems again from the differences in their dynamical variables.

We now try to find the gravitational field inside and outside of a sphere of radius R filled with a dust where only the temporal component of $T_{\mu i}$ is non-zero, $T_{00}=-\rho$. We also assume that the density, $\rho$, is constant and locate the origin of the coordinate system at the center of the sphere.
The result can be reached by directly solving field equation (\ref{linearField5})
\bqn
\lb{linearfield40}
\nabla ^2 A_{i00}= 4\pi G y_i\partial_i\rho(x).
\eqn
The field should be zero at infinity and should not diverge at the center. To find the boundary condition at $r=R$, one needs to integrate equation (\ref{linearfield40})
\bqn
\int d^3x \nabla ^2 A_{i00} = 4\pi G \int d^3x y_i\partial_i\rho.
\eqn
Using Gauss' theorem and knowing that $\vec{A}\equiv(0,A_{100},A_{200},A_{300})$ is radial, the integral reduces to
\bqn
\lb{boundarycondition}
\partial_r \left(\vec{A}\cdot\hat{r}\right)\Bigr|_{R^+}-\partial_r \left(\vec{A}\cdot\hat{r}\right)\Bigr|_{R^-} =-4\pi G \rho,
\eqn
which is the last piece of the boundary conditions. Now, we need to solve field equation (\ref{linearfield40}) knowing that, except at the boundary $r=R$, the source is everywhere zero 
\bqn
\nabla ^2 \vec{A}=0.
\eqn
It is now easy to verify that  
\bqn
A_{i 0 0}=
\lb{linearField10}
\begin{cases}
\frac{MG\vec{x}_i}{|\vec{x}|^3}y_i &  |\vec{x}| > R,\\
\frac{MG\vec{x}_i}{R^3}y_i &  |\vec{x} | < R.
\end{cases}
\eqn
satisfies both the field equation and the boundary conditions.

To find the Newtonian gravitational force, we start with the geodesic equation, $D_{\mu}\dot{x}^i=0$, or consequently
\bq
\ddot{x}^i-A^i_{~j \mu}\dot{x}^j\dot{x}^{\mu}=0.
\eq
After neglecting the spatial components of velocity in the lowest order of perturbation, and inserting $\dot{x}^{\mu=0}=-1$ and $\dot{x}^{j=0}=1$
\bq
\ddot{x}^i=-A^i_{~00}.
\eq
This can be written in the final form
\bq
\vec{\ddot{x}}=\begin{cases}
-\frac{MG\vec{x}}{|\vec{x}|^3} &  |\vec{x}| > R,\\
-\frac{MG\vec{x}}{R^3} &  |\vec{x}| < R.
\end{cases}.
\eq
This shows that both the interior and exterior Schwarzschild space-times are the linear order solutions of the present theory.

\section{Effective Lorentz Gauge Theory: General Non-Linear Case}
The scope 
of this section is to find an effective Lorentz gauge theory in a general non-linear case. As was explained in the previous section, due to the fluctuations in the source field, it is extremely hard to use Lorentz gauge theory in describing classical configurations. Therefore, instead of using equation \eqref{SimpleField} in macroscopic problems, we would like to utilize the effective theory whose linear version was defined in the preceding section. In order to find the non-linear format, we note that in a free falling frame, the field equations of the generally non-linear effective theory is linear and is given by equation \eqref{linearField5}. A simple frame transformation will give the field equations in an arbitrary frame. This is equivalent to replacing the partial derivatives in \eqref{linearField5} by covariant derivatives \cite{Will2006}. Therefore, 
the effective field equations can be written as 
\bqn
\lb{classicalFieldEq}
D_{\nu}F^{\mu\nu ij}&=&4\pi G \tilde{J}^{\mu ij},
\eqn
where the effective source is
\bqn
\lb{generalsource}
\tilde{J}_{\mu ij}=D_j T_{\mu i}-D_i T_{\mu j}.
\eqn
The conservation law of the classical source implies that
\bqn
D_{\mu}\tilde{J}^{\mu ij}=0,
\eqn
since the left hand side of (\ref{classicalFieldEq}) is divergence free, $D_{\mu}D_{\nu}F^{\mu\nu ij}=0$.

\subsection{Interior Of A Spherically Symmetric Star}
In the following we use the general theory to find a solution for spherical static incompressible stars, corresponding with the interior Schwarzschild solution in GR.
A spherically symmetric space-time is represented by the following tetrad
\bqn
e_{i \mu}=
\begin{pmatrix}
\sqrt{a(r)}&~&~&~\\
~&\sqrt{b(r)}&~&~\\
~&~&r&~\\
~&~&~&r\sin(\theta)
\end{pmatrix}.
\eqn
In appendix A a Mathematica code is provided that computes $\Gamma^{\alpha}_{\mu\nu}$, $A_{ij\mu}$, $F_{\mu\nu ij}$, and field equations as well as the source field, its divergence and the divergence of energy-momentum and the results are presented below. The Christoffel symbols, $\Gamma^{\lambda}_{\mu\nu}$, can be easily calculated
using (\ref{Chris})
\begin{align}
\Gamma^1_{00}&=\frac{a'}{2b} , & \Gamma^2_{12}&=\frac{1}{r},& \Gamma^1_{22}&=-\frac{r}{b}, \nb\\
\Gamma^0_{01}&=\frac{a'}{2a}, & \Gamma^3_{13}&=\frac{1}{r},& \Gamma^1_{33}&=-\frac{r\sin^2(\theta)}{b}, \nb\\
\Gamma^1_{11}&=\frac{b'}{2b}, & \Gamma^3_{23}&=\frac{\cos(\theta)}{\sin(\theta)},& \Gamma^2_{33}&=-\sin(\theta)\cos(\theta), \nb\\
\end{align}
where prime indicates derivative with respect to r. The spin connections, $A_{ij\mu}$, using (\ref{SpinCon}) are
\begin{align}
A_{100}&=\frac{a'}{2\sqrt{ab}},&  A_{122}&=\frac{1}{\sqrt{b}},\nb\\
A_{133}&=\frac{\sin(\theta)}{\sqrt{b}},& A_{233}&=\cos(\theta),\nb\\
\end{align}
and the strength tensor, $F_{\mu\nu ij}$, using (\ref{strength}) reads
\begin{align}
F_{1010}&=-\frac{a''}{2\sqrt{ab}}+\frac{a'b'}{4\sqrt{ab^3}}+\frac{a'^2}{4\sqrt{a^3b}} , &
F_{0220}&=\frac{a'}{2\sqrt{a}b}, & F_{0330}&=\frac{\sin(\theta)a'}{2\sqrt{a}b},\nb\\
F_{1221}&=-\frac{b'}{2\sqrt{b^3}}, & F_{1331}&=-\frac{\sin(\theta)b'}{2\sqrt{b^3}}, &
 F_{3232}&=\sin(\theta)\big(1-\frac{1}{b}\big).
\end{align} 
Here, and also in the rest of the paper, only nonzero components are shown.
The energy-momentum tensor is given by
\bqn
\lb{energymomentum}
T_{\mu}^{~\nu}=
\begin{pmatrix}
-\rho&~&~&~\\
~&p&~&~\\
~&~&p&~\\
~&~&~&p
\end{pmatrix},
\eqn
and therefore the independent and non-zero components of $\tilde{J}^{\mu ij}$ are
\bqn
&&\tilde{J}^{\theta \theta r}=\frac{p'}{r\sqrt{b}},\nb\\
&&\tilde{J}^{t t r}=-\frac{\rho '}{\sqrt{ab}}-\frac{a' (\rho+p)}{2\sqrt{a^3 b}}.
\eqn
Substituting all the pieces into (\ref{classicalFieldEq}) gives the following two independent equations 
\bqn
\lb{interiorfieldequations}
\frac{-1}{4r^2\sqrt{(ab)^7}} \Bigg(&&2r^2abb'a'^2 - 3 r^2 ba^2a''b'-r^2ba^2a'b''+2r^2a^2b^2a'''+2r^2a^2a'b'^2+2r^2b^2a'^3-4r^2ab^2a'a''\nb\\
&&-2rab^2a'^2-2ra^2ba'b'+4ra^2b^2a''-4a^2b^2a' ~~\Bigg)=4\pi G \Bigg(-\frac{\rho '}{\sqrt{ab}}-\frac{a' (\rho+p)}{2\sqrt{a^3 b}}\Bigg),\nb\\
~~\nb\\
\frac{1}{4r^4a^2\sqrt{b^7}}\Bigg(&& r^2 b^2a'^2+r^2aba'b'-4r^2a^2b'^2+2r^2a^2bb''-4a^2b^3+4a^2b^2 ~~\Bigg)=4\pi G \frac{p'}{r\sqrt{b}}.
\eqn
Conservation of energy-momentum tensor implies that, see \cite{adler1965introduction},
\bqn
\lb{energymomentumconservation}
p'=-\frac{(\rho+p)a'}{2a}.
\eqn
A direct substitution shows that $D_{\mu}\tilde{J}^{\mu ij}=0$ is trivially satisfied with no further constraint. 

All the necessary differential equations for investigating spherically symmetric problems are now available. In \cite{Borzou2016} it is shown that the exterior Schwarzschild space-time is an exact solution to the equations above. Also, through equation (\ref{linearField10}) we showed that the interior Schwarzschild space-time is a first order solution of the equations as well. To find out the non-linear solution to the incompressible star, a numerical study will be performed. But before moving to the numerical study, it is worth mentioning that the exact interior Schwarzschild space-time satisfies the differential equations if both density and pressure were negative. This can be verified by first flipping the sign of the right hand sides of (\ref{interiorfieldequations}) and then directly substituting the followings 
\bqn
\lb{interiorSchwarzschild}
&&a=\Bigg(\frac{3}{2}\sqrt{1 - \frac{2GM}{R}} - \frac{1}{2}\sqrt{1 - \frac{2GM}{R^3}r^2}\Bigg)^2,\nb\\
&&b=\frac{1}{1-\frac{2GM}{R^3}r^2},\nb\\
&&p=\rho\Bigg( \frac{\sqrt{1 - \frac{2GM}{R}} - \sqrt{1 - \frac{2GM}{R^3}r^2}}
{\sqrt{1 - \frac{2GM}{R^3}r^2}-3\sqrt{1 - \frac{2GM}{R}}} \Bigg),
\eqn
where $\rho$ is a positive constant.

\subsection{A Numerical Solution to the Interior of an Incompressible Star}
\begin{figure}[tbp]
\centering
\includegraphics[width=5.5cm]{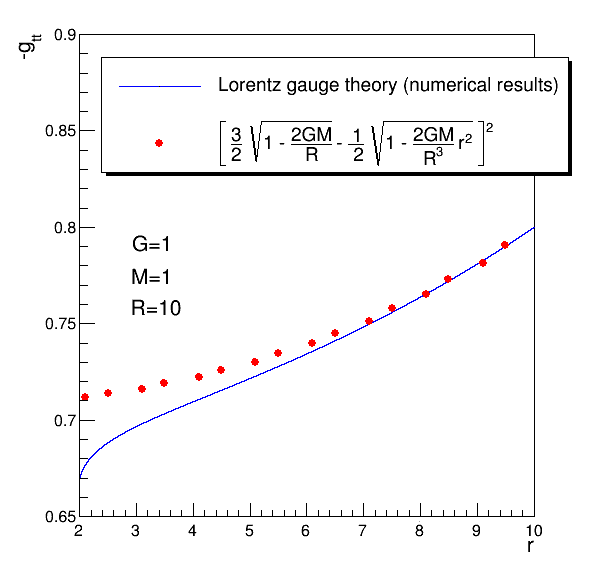}
\includegraphics[width=5.5cm]{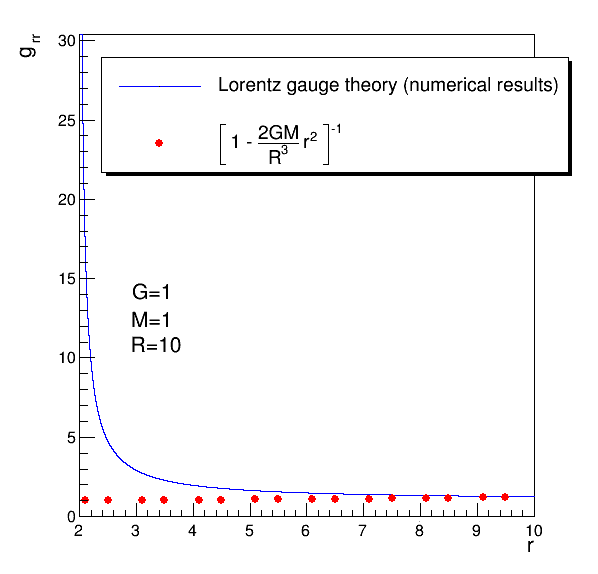}
\includegraphics[width=5.5cm]{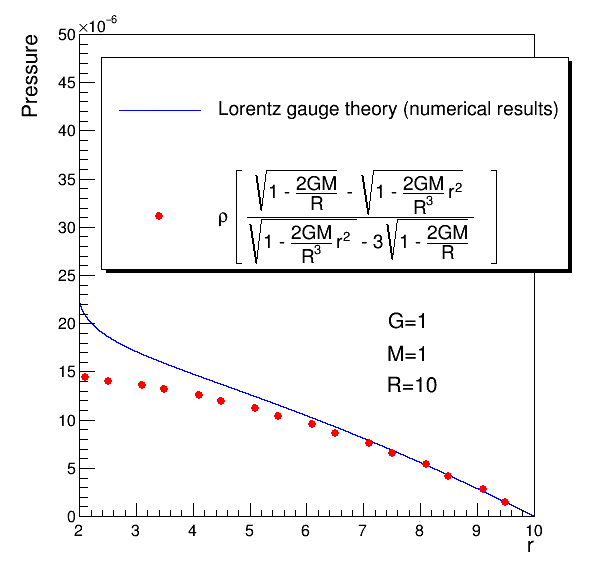}
\caption{Numerical results from the Lorentz gauge theory versus the corresponding analytic solutions from general relativity. (Left) The temporal component of the metric (Middle) the radial component of the metric (Right) the pressure inside an incompressible star in a static spherically symmetric space-time.}
\lb{numericalSolutionFig}
\vspace{0.75cm}
\end{figure} 
In this section we would like to numerically solve equations (\ref{interiorfieldequations}) and (\ref{energymomentumconservation}) using the simplest Runge-Kutta method. We start with the set of equations in (\ref{interiorfieldequations}) and by rearranging them find an explicit expression for  $a'''$ and $b''$. Equation (\ref{energymomentumconservation}) is already in the desired format. To find the initial conditions we use the fact that $a$, $a'$, $b$, $b'$, and $p$ are all continuous at the surface of the star $r=R$ and use equation (\ref{boundarycondition}) to find a value for $a''$ given that the exterior of the star is governed by the exact Schwarzschild space-time. 
For simplicity we also set
\bqn
&&G=M=1,\nb\\
&&R=10.
\eqn 
Also density is a constant in terms of $R$ and $M$. The initial values at the surface are given in table \ref{tableinitialvalue}.
\begin{table}[th]
\caption{
Initial conditions used for numerically solving the field equation.
}
\begin{center}
\begin{tabular}{c|cccccc}
\hline
Quantity       & ~~$p$ & ~~$a$~~ & $a'$~ & $a''$ ~ & $b$  ~& $b'$   \\
\hline
Value at $r=R$ & ~~0   & ~~0.8~~ & 0.02~ & 0.002 ~ & 1.25 ~& -0.03125  \\
\hline
\end{tabular}
\end{center}
\label{tableinitialvalue}
\end{table}

Values of $a'''$, $b''$, and $p'$ are also known at the surface through equations (\ref{interiorfieldequations}) and (\ref{energymomentumconservation}). To determine the values of the quantities at one step away from the surface, the following equation can be used
\bqn
f(r-\Delta r)&=&f(r)-\Delta r f'(r),
\eqn
where $f$ stands for $p$, $a$, $a'$, $a''$, $b$, and $b'$.
Knowing the value of these quantities at $R - \Delta r$, one can find the values of $a'''$, $b''$, and $p'$ at the same point through the equations mentioned above. Now we can go further back and find the values at $R - 2\Delta r$ and beyond by simply repeating the same procedure. We have programmed this algorithm in C++ and plotted using ROOT \cite{Brun:1997pa}. The results are shown in figure \ref{numericalSolutionFig} where the solid line represents the numerical values and the points show the values of equation (\ref{interiorSchwarzschild}) at a few occasional locations.  The agreement between the solid lines and the dots from the surface to $r\sim 6$ is consistent with our expectations from equation (\ref{linearField10}). Also there seems to exist a singularity at $r\sim 2$, for which the plots are cut at this point. Study of this singularity is out of the scope of the present article and is left for future studies.

\section{Conclusions}

An effective Lorentz gauge theory of gravity for macroscopic scales have been derived.
The effective theory was necessary because the exact source field in equation (\ref{SimpleField}) is more convenient for microscopic problems. By investigating the first order of perturbation of the underlying equations, we have found a linear effective theory that works well provided the gravitational fields are negligible in extremely small distances. Next, using the principle of minimal coupling, we have derived the non-linear formulation of the effective theory and have shown that it takes the following form
\bqn
D_{\nu}F^{\mu\nu ij}&=&4\pi G \tilde{J}^{\mu ij}\nb\\
&=&4\pi G \left(D^j T^{\mu i}-D^i T^{\mu j}\right),
\eqn
where $T^{\mu i}$ is the energy-momentum tensor, and $G$ is Newton's gravitational constant and is emerged by integrating a small length out of the underlying equations. 
This equation has been used to investigate the interior of an incompressible star. Through numerical studies it is shown that the effective theory within the star is consistent with the interior Schwarzschild space-time in the linear regions but deviates from it in the non-linear regions.

It also has been shown that bosonic fields do not contribute to the energy-momentum tensor defined as 
\bqn
T^{\mu i}=\frac{\delta {\cal{L}}}{\delta e_{i \mu}},
\eqn
due to the non-dynamical nature of the metric and the fact that the tetrad field does not directly contribute to the bosonic Lagrangians. Nevertheless, bosons are still affected by gravity through the metric that is coupled to their Lagrangians and light, for example, does bend under the gravitational fields. The difference with GR is that bosonic fields do not gravitate with the consequence that even a charged spherical black hole generates the Schwarzschild, rather than the Reissner-Nordström, metric.

~\\{\bf Acknowledgements:}
We are grateful to professors Gerald Cleaver and Hamid Reza Sepangi for careful reading
of the draft and useful suggestions. Our special thanks are
due to Jay Dittmann and Kenichi Hatakeyama for their continued support. In the end we would like to express our gratitude to the referees whose comments and suggestions significantly added to the present work. 

\section*{Appendix A: Mathematica Code}
The following pieces of Mathematica code, each enclosed by two lines, are developed to perform the calculations presented in this paper. By pasting them into a Mathematica notebook, one should be able to reproduce the results. We first would like to emphasize that in Mathematica the indices run from 1 and not 0, therefore the temporal components are moved to the fourth place. The code starts with the following piece
\bq
\begin{tabular}{l}
\hline
\\
Clear[ee, e, eta, eeta, g, ggInv, gInv, coord,M, R, rho, p, a, b]     \\
theta = $\theta$;\\
phi=$\phi$;\\
\\
\hline
\end{tabular}\nb
\eq
Next the coordinates are defined
\bq
\begin{tabular}{l}
\hline
\\
coord = \{r, theta, phi, t\};    \\
\\
\hline
\end{tabular}\nb
\eq

Define the tetrad, $e_{i\mu}$ such that $a$ and $b$ are only a function of r
\bq
\begin{tabular}{l}
\hline
\\
ee = \{\{Sqrt[b[r]], 0, 0, 0\}, \{0, r, 0, 0\},
\{0, 0, r * Sin[theta], 0\}, \{0, 0, 0, Sqrt[a[r]]\}\};    \\
e[i\_, mu\_] := e[i, mu] = ee[[i, mu]];\\
\\
\hline
\end{tabular}\nb
\eq

~\\
Both $\eta _{i j}$ and $\eta ^{i j}$, flat metric, are
 \bq
\begin{tabular}{l}
\hline
\\
eeta = \{\{1, 0, 0, 0\}, \{0, 1, 0, 0\}, \{0, 0, 1, 0\}, \{0, 0, 0, -1\}\};\\
eta[i\_, j\_] := eta[i, j] = eeta[[i, j]];\\
\\
\hline
\end{tabular}\nb
\eq

~\\
Next the metric, $g_{\mu \nu}$, and its inverse,  $g^{\mu \nu}$, are introduced
 \bq
\begin{tabular}{l}
\hline
\\
g[mu\_, nu\_] := g[mu, nu] = $\sum_{i=1}^{4}\limits\sum_{j=1}^{4}\limits$ eta[i, j] * e[i, mu] * e[j, nu]; \\
ggInv = Inverse[Table[g[mu, nu], \{mu, 1, 4\}, \{nu, 1, 4\}]]; \\
gInv[mu\_, nu\_] := gInv[mu, nu] = ggInv[[mu, nu]];\\
\\
\hline
\end{tabular}\nb
\eq

~\\
The Christoffel symbols, $\Gamma^{\mu}_{\nu\alpha}$
 \bq
\begin{tabular}{l}
\hline
\\
Clear[christ]\\
christ[a\_, b\_, c\_] := christ[a, b, c] = Simplify[\\
~~~~$\sum_{d=1}^{4}\limits$1/2* gInv[a, d]*(D[g[d, c], coord[[b]]] + D[g[d, b], coord[[c]]] - D[g[b, c], coord[[d]]])\\
];\\
\\
\hline
\end{tabular}\nb
\eq
where the non-zero components can be displayed by 
\bq
\begin{tabular}{l}
\hline
\\
TableForm[\\
~~DeleteCases[\\
~~~~Flatten[\\
~~~~~~Table[\\
~~~~~~~~If[UnsameQ[christ[mu, nu, alpha], 0],\\
~~~~~~~~\{ToString[$\Gamma$[coord[[mu]], coord[[nu]], coord[[alpha]]]] -$>$
christ[mu, nu, alpha]\}]\\
~~~~~~~~, \{mu, 1, 4\}, \{nu, 1, 4\}, \{alpha, 1, 4\}\\
~~~~~~], 2\\
~~~~], Null\\
~~]\\
]\\
\\
\hline
\end{tabular}\nb
\eq
which shows a group of terms like $\Gamma[r,\phi,\phi]\rightarrow -\frac{r Sin[\theta]^2}{b[r]}$.

~\\
Next the gauge fields, $A_{i j \mu}$, are calculated
 \bq
\begin{tabular}{l}
\hline
\\
Clear[A]\\
A[i\_, j\_, mu\_] := A[i, j, mu] =$\sum_{\text{nu}=1}^{4}\limits$\Bigg(\\
~~~~\Big(~$\sum_{\text{beta}=1}^{4}\limits$gInv[nu, beta] * e[j, beta]\Big)*\\
~~~~\Big(D[e[i, nu], coord[[mu]]] -\big(~$\sum_{\text{alpha}=1}^{4}\limits$christ[alpha, mu, nu] * e[i, alpha]\big)\Big)\\
\Bigg);\\
\\
\hline
\end{tabular}\nb
\eq
The non-zero components can be displayed by
\bq
\begin{tabular}{l}
\hline
\\
TableForm[\\
~~DeleteCases[\\
~~~~Flatten[\\
~~~~~~Table[\\
~~~~~~~~If[UnsameQ[A[mu, nu, alpha], 0],\\
~~~~~~~~\{ToString[AA[coord[[mu]], coord[[nu]], coord[[alpha]]]] -$>$ A[mu, nu, alpha]\}],
\{mu, 1, 4\}, \{nu, 1, 4\}, \{alpha, 1, 4\}\\
~~~~~~], 2\\
~~~~], Null\\
~~]\\
]\\
\\
\hline
\end{tabular}\nb
\eq

~\\
Next we can calculate and display the strength tensor, $F_{\mu\nu ij}$, by
\bq
\begin{tabular}{l}
\hline
\\
Clear[F]\\
F[mu\_, nu\_, i\_, j\_] := F[mu, nu, i, j] =\\
~~~~D[A[i, j, mu], coord[[nu]]] + $\sum_{m=1}^{4}\limits\sum_{n=1}^{4}\limits$eta[m, n] * A[i, n, mu] * A[m, j, nu]\\
~~~-D[A[i, j, nu], coord[[mu]]] - $\sum_{m=1}^{4}\limits\sum_{n=1}^{4}\limits$eta[m, n] * A[i, n, nu] * A[m, j, mu];\\
~\\
TableForm[\\
~~DeleteCases[\\
~~~~Flatten[\\
~~~~~~Table[\\
~~~~~~~~If[UnsameQ[F[mu, nu, i, j], 0],\\
~~~~~~~~ \{ToString[FF[coord[[mu]],
coord[[nu]], coord[[i]], coord[[j]]]] -$>$ F[mu, nu, i, j]\}],\\
~~~~~~~~\{mu, 1, 4\}, \{nu, 1, 4\}, \{i, 1, 4\}, \{j, 1, 4\}\\
~~~~~~], 3\\
~~~~], Null\\
~~]\\
]\\
\\
\hline
\end{tabular}\nb
\eq

~\\
The right hand side of the field equations, $D_{\nu}F^{\mu \nu i j}$, is calculated by
\bq
\begin{tabular}{l}
\hline
\\
Clear[field]\\
field[mumu\_, ii\_, jj\_] := field[mumu, ii, jj] = 
$\sum_{\text{mu}=1}^{4}\limits\sum_{i=1}^{4}\limits\sum_{j=1}^{4}\limits$ gInv[mumu, mu] * eta[ii, i] * eta[jj, j] *\Bigg(\\
~~$\sum_{\text{alpha}=1}^{4}\limits\sum_{\text{nu}=1}^{4}\limits$ gInv[alpha, nu] * \Big(\\
~~~~~~D[F[mu, nu, i, j], coord[[alpha]]]\\
~~~~~~ -$\sum_{\text{beta}=1}^{4}\limits$ christ[beta, alpha, mu] * F[beta, nu, i, j]\\
~~~~~~ -$\sum_{\text{beta}=1}^{4}\limits$ christ[beta, alpha, nu] * F[mu, beta, i, j]\\
~~~~~~ -$\sum_{m=1}^{4}\limits\sum_{n=1}^{4}\limits$eta[m, n] * A[i, m, alpha] * F[mu, nu, n, j]\\
~~~~~~ -$\sum_{m=1}^{4}\limits\sum_{n=1}^{4}\limits$eta[m, n] * A[j, m, alpha] * F[mu, nu, i, n]\\
~~\Big)\\
\Bigg)\\
\\
\hline
\end{tabular}\nb
\eq
To display terms like $D_{\nu}F^{t \nu t r}$, one can run
\bq
\begin{tabular}{l}
\hline
\\
Expand[field[4, 4, 1]]\\
\\
\hline
\end{tabular}\nb
\eq

~\\
The energy-momentum tensor, $T^{\mu \nu}$, and its divergence are introduced, calculated, and displayed by
\bq
\begin{tabular}{l}
\hline
\\
Clear[Tupdown, T, DT]\\
Tupdown = \{\{p[r], 0, 0, 0\}, \{0, p[r], 0, 0\}, \{0, 0, p[r], 0\}, \{0, 0, 0, -rho[r]\}\};\\
T[mu\_, nu\_] := T[mu, nu] = $\sum_{\text{alp}=1}^{4}\limits$ gInv[mu, alp] * Tupdown[[alp, nu]];\\
DT[nu\_] := DT[nu] = $\sum_{\text{mu}=1}^{4}\limits$\Big(\\
~~D[T[mu, nu], coord[[mu]]] + $\sum_{\text{alpha}=1}^{4}\limits$ (christ[mu, mu, alpha] * T[alpha, nu] +
christ[nu, mu, alpha] * T[mu, alpha])\\
\Big);\\
~~\\
TableForm[\\
~~DeleteCases[\\
~~~~Flatten[\\
~~~~~~Table[\\
~~~~~~~~If[UnsameQ[DT[mu, nu], 0], \{ToString[DTDT[coord[[nu]]]] -$>$ DT[nu]\}],
\{nu, 1, 4\}\\
~~~~~~], 0\\
~~~~], Null\\
~~]\\
]\\
\\
\hline
\end{tabular}\nb
\eq

~\\
The source field, $J^{\mu i j}=D^jT^{\mu i}-D^iT^{\mu j}$, and its divergence, $D_{\mu}J^{\mu i j}$, can be calculated by
\bq
\begin{tabular}{l}
\hline
\\
Clear[J,DJ]\\
J[mu\_, i\_, j\_] := J[mu, i, j] =\\
~~$\sum_{\text{nu}=1}^{4}\limits\sum_{\text{lambda}=1}^{4}\limits$\Big(\\
~~~~$\sum_{\text{alpha}=1}^{4}\limits\sum_{m=1}^{4}\limits\sum_{n=1}^{4}\limits$\big(\\
~~~~~~~~~~eta[j, m] * gInv[nu, alpha] * e[m, alpha] *
eta[i, n] * e[n, lambda]\\
~~~~~~~~~~ - eta[i, m] * gInv[nu, alpha] *
e[m, alpha] * eta[j, n] * e[n, lambda]\\
~~~~\big)\\
~~\Big)*\Big(\\
~~~~D[T[mu, lambda], coord[[nu]]] +\\
~~~~ $\sum_{beta=1}^{4}\limits$ \big(
christ[mu, nu, beta] * T[beta, lambda] +
christ[lambda, nu, beta] * T[mu, beta] \big)\\
~~\Big);\\
~~\\
TableForm[\\
~~DeleteCases[\\
~~~~Flatten[\\
~~~~~~Table[\\
~~~~~~~~If[UnsameQ[J[mu, nu, alpha], 0],\\
~~~~~~~~\{ToString[JJ[coord[[mu]], coord[[nu]], coord[[alpha]]]] -$>$ J[mu, nu, alpha]\}],
\{mu, 1, 4\}, \{nu, 1, 4\}, \{alpha, 1, 4\}\\
~~~~~~], 2\\
~~~~], Null\\
~~]\\
]\\
~~\\
DJ[i\_, j\_] := DJ[i, j] = $\sum_{\text{mu}=1}^{4}\limits$\Bigg(\\
~~~~D[J[mu, i, j], coord[[mu]]] +\\
~~~~$\sum_{\text{alpha}=1}^{4}\limits$\big(christ[mu, mu, alpha] * J[alpha, i, j]\big)-\\
~~~~$\sum_{m=1}^{4}\limits\sum_{n=1}^{4}\limits$\big( eta[i, n] * A[n, m, mu] * J[mu, m, j] +
eta[j, n] * A[n, m, mu] * J[mu, i, m] \big)\\
\Bigg);\\
~~\\
Table[Simplify[DJ[i, j]], \{i, 1, 4\}, \{j, 1, 4\}]\\
\\
\hline
\end{tabular}\nb
\eq


\newpage
\begin{thebibliography}{nbound}
\bibitem{Weinberg:1980gg} S. Weinberg, Ultraviolet Divergences in Quantum Theories of Gravitation, in: General
Relativity. An Einstein Centenary Survey, Cambridge U. P. (1980). eds: S. W. Hawking and W. Israel. 

\bibitem{PhysRev.101.1597} R. Utiyama, Phys. Rev. 101 (1956) 1597.

\bibitem{RevModPhys.36.463} D. W. Sciama, Rev. Mod. Phys. 36 (1964) 463.

\bibitem{Kibble1961} T. W. B. Kibble, J. Math. Phys. 2 (1961) 212. 

\bibitem{Borzou2016} A. Borzou, Class. Quantum Grav. 33 (2016) 025008. [arXiv:1412.1199]

\bibitem{Burgess:2007} C. P. Burgess, Ann. Rev. Nucl. Part. Sci. 57 (2007) 329. [arXiv:hep-th/0701053]

\bibitem{Bain:2013imf} J. Bain, Effective field theories, in: The Oxford Handbook of Philosophy of Physics, Oxford University Press (2013). eds: R. Batterman

\bibitem{GEORGI1993} H. Georgi, Ann. Rev. Nucl. Part. Sci. 43 (1993) 209.

\bibitem{Will2006} C. M. Will, Living Rev. Relativity 17 (2014) 4. [arXiv:1403.7377]

\bibitem{Hehl19951} F. W. Hehl, J. D. McCrea, E. W. Mielke and Y. Neeman, Phys. Rep. 258 (1995) 1. 
\bibitem{Blagojevic:2013xpa} M. Blagojević and F.~W. Hehl, \newblock {\em {Gauge Theories of Gravitation}}, \newblock World Scientific, Singapore (2013).

\bibitem{IVANENKO19831} D. Ivanenko and G. Sardanashvily, Phys. Rep. 94 (1983) 1. 
\bibitem{RevModPhys.48.393} F. W. Hehl, P. von~der Heyde, G. D. Kerlick, and J. M. Nester, \newblock {\em Rev. Mod. Phys.} 48 (1976) 393.
\bibitem{Obukhov:2006gea} Y. N. Obukhov, \newblock {\em Int. J. Geom. Meth. Mod. Phys.} 3 (2006) 95. [arXiv:gr-qc/0601090]
\bibitem{Shapiro2002113} I. L. Shapiro, \newblock {\em Physics Reports} 357 (2002) 113. [arXiv:hep-th/0103093]

\bibitem{Hamber2009} H. W. Hamber, Quantum  Gravitation, Springer  Tracts  in  Modern  Physics, New York (2009). 
\bibitem{kiefer2012} C. Kiefer, \newblock {\em Quantum Gravity: Third Edition}, \newblock International Series of Monographs on Physics. OUP Oxford (2012).

\bibitem{Isham1991} C. J. Isham, \newblock {\em Recent Aspects of Quantum Fields: Proceedings of the XXX Int.
  Universit{\"a}tswochen f{\"u}r Kernphysik, Schladming, Austria February and
  March 1991}, chapter Conceptual and geometrical problems in quantum gravity, \newblock Springer, Berlin Heidelberg (1991) 123.

\bibitem{Casimir1948} H. B. G. Casimir, Proc. K. Ned. Akad. Wet. 51 (1948) 793.
\bibitem{Sparnaay1957} M. J. Sparnaay, Nature 180 (1957) 334.
\bibitem{RevModPhys.61.1} S. Weinberg, \newblock {\em Rev. Mod. Phys.} 61 (1989) 1. [arXiv:astro-ph/0005265]
\bibitem{Baltz:2004tj} E. A. Baltz, \newblock {Dark matter candidates}, 32nd SLAC summer institute on particle physics: Natures Greatest Puzzles Menlo Park, California (2004).

\bibitem{Hayashi01091980} K. Hayashi and T. Shirafuji, Prog. Theor. Phys. 64 (1980) 866.
\bibitem{PhysRevD.80.104031} V. P. Nair, S. Randjbar-Daemi, and V. Rubakov, Phys. Rev. D 80 (2009) 104031. [arXiv:hep-th/0811.3781]

\bibitem{VonDerHeyde1975} P.~von der~Heyde, \newblock {\em Lettere al Nuovo Cimento} 14 (1975) 250.

\bibitem{Obukhov1989} Yu.~N. Obukhov, V.~N. Ponomariev, and V.~V. Zhytnikov, \newblock {\em General Relativity and Gravitation} 21 (1989) 1107.

\bibitem{BAEKLER1987800} P.~Baekler, E. W. Mielke, R.~Hecht, and F. W. Hehl, \newblock {\em Nuclear Physics B} 288 (1987) 800.

\bibitem{Vereshchagin:1999jv} G.~V. Vereshchagin, A.~S. Garkun, and A.~V. Minkevich, \newblock {\em Grav. Cosmol. Suppl.} 5 (1999) 6.

\bibitem{Kuhfuss1986} R. Kuhfuss and J. Nitsch, \newblock {\em General Relativity and Gravitation} 18 (1986) 1207.

\bibitem{ASNA:ASNA2103020310} F. W. Hehl, J. Nitsch, and P. von der Heyde, Gravitation and the Poincare gauge field theory with quadratic Lagrangian, in: General Relativity and Gravitation: One hundred years after the birth of Albert Einstein, 
Plenum Press, New York \& London (1980). eds: A. Held.

\bibitem{Weinberg1972} S. Weinberg, Gravitation And Cosmology: Principles And Applications Of The General Theory Of Relativity, 
John Wiley and Sons (1972).

\bibitem{SIJACKI1982435} Dj. Sijacki, \newblock {\em Physics Letters B} 109 (1982) 435. 

\bibitem{adler1965introduction} R. Adler, M. Bazin, and M. Schiffer, Introduction to General Relativity, McGraw-Hill, New York (1975).
\bibitem{Brun:1997pa} R. Brun and F. Rademakers, 
ROOT: An Object Oriented Data Analysis Framework, 
Proceedings AIHENP'96 Workshop, Lausanne, Sep. 1996, Nucl. Inst. \& Meth. in Phys. Res. A 389 (1997) 81. See also http://root.cern.ch/.



\end{thebibliography}

\end{document}